\documentclass[pra,superscriptaddress,showpacs,twocolumn,aps,10pt]{revtex4-1}

\usepackage{amsfonts,amssymb,amsmath,color}
\usepackage{graphics,graphicx,epsfig}
\usepackage{epstopdf}

\newcommand{\be}{\begin{eqnarray}}
\newcommand{\ee}{\end{eqnarray}}

\newcommand{\ba}{\begin{aligned}}
\newcommand{\ea}{\end{aligned}}

\usepackage{graphicx}
\begin{document}

\title{Experimental Test of Nonclassicality with Arbitrary Low Detection Efficiency}

\author{Alley Hameedi}
\affiliation{Department of Physics, Stockholm University, S-10691 Stockholm, Sweden}

\author{Breno Marques}
\affiliation{Centro de Ci\^encias Naturais e Humanas, Universidade Federal do ABC - UFABC, Santo Andr\'e, Brazil}

\author{Piotr Mironowicz}
\affiliation{Department of Algorithms and System Modeling, Faculty of Electronics, Telecommunications and Informatics, Gda\'{n}sk University of Technology, Gda\'{n}sk 80-233, Poland}
\affiliation{International Centre for Theory of Quantum Technologies, University of Gda\'nsk, Wita Stwosza 57, 80-308 Gda\'nsk, Poland}

\author{Debashis Saha}
\affiliation{Institute of Theoretical Physics and Astrophysics, University of Gda\'{n}sk, 80-952 Gda\'{n}sk, Poland}

\author{Marcin Paw\l{}owski}
\email{dokmpa@univ.gda.pl}
\affiliation{Institute of Theoretical Physics and Astrophysics, University of Gda\'{n}sk, 80-952 Gda\'{n}sk, Poland}
\affiliation{International Centre for Theory of Quantum Technologies, University of Gda\'nsk, Wita Stwosza 57, 80-308 Gda\'nsk, Poland}

\author{Mohamed Bourennane}
\affiliation{Department of Physics, Stockholm University, S-10691 Stockholm, Sweden}

\date{\today}

\begin{abstract}
We theoretically introduce and experimentally demonstrate the realization of a nonclassicality test that allows for arbitrarily low detection efficiency without invoking an extra assumptions of independence of the devices. Our test and its implementation is set in a prepare-and-measure scenario with an upper limit on the classical communication capacity of the channel through which the systems are communicated. The essence for our novel test is the use of two preparation and two measurement devices, which are randomly paired in each round. Our work opens the possibility of experimental realizations of nonclassicality tests with off-the-shelf technology.
\end{abstract}

\maketitle

\section{Introduction}

Quantum Information processing utilizes nonclassical resources for processing information encoded into physical systems, which are subjected to quantum mechanical laws. These quantum resources have many applications in quantum information processing ranging from cryptography~\cite{BB84,E91} to communication complexity reduction~\cite{BCD}.

Device independent quantum information processing schemes have shown great promise in the estimation of critical parameters, without making any assumptions about the inner functionality of these devices. This approach comes in two main variants: fully-device-independent (FDI)~\cite{DI} and semi-device independent (SDI)~\cite{SDI}. Fully-device-independent protocols means that the working principles of the devices does not matter and the protocols works without assuming any internal mechanism of the devices. On the other hands, SDI protocol means that the protocols requires some assumptions on the devices in order to be realized. In this paper we consider the SDI approach. The assumption we make is an upper bound on the classical communication between parts of devices involved in the scenario.

No control over the communicated system renders tasks as cryptography and randomness generation impossible. Therefore, we suppose that we can find at least one parameter, which describes the communication: an upper bound on the classical capacity of the channel. This assumption is the reason why the scenario is termed semi-device-independent.

The implementation of FDI or SDI protocols requires a test of nonclassicality to be performed, a result of which indicates that the description of the experiment is impossible through classical means. This is a necessary, although not always a sufficient condition. These tests include estimation of violations of Bell inequality~\cite{bell}, dimension witnesses~\cite{dimwit} or success probability in a communication complexity task~\cite{M-expDW}. The important requirement, however, is that the involved parties must give a conclusive result without any additional assumptions. It is difficult in practice, mainly due to the so-called detection efficiency loophole~\cite{Pearle}, which in a nutshell states that device independent tests of nonclassicality can only be conclusive if the detectors used in an experiment provide a detection efficiency above a certain threshold. These thresholds have been found for a variety of tests~\cite{eberhard,serge,cabello,eff,rob,M-PBR}. Unfortunately, the required efficiencies are very high and therefore difficult to obtain in practice. To our knowledge only a few DI tests of nonclassicality have been performed thus far~\cite{di2,di3,di4,di5,di6}.

The issue can be addressed by adding extra constraints on the resources that are available to the devices. One such constraint is the requirement that the parties have no access to shared randomness. With this added assumption, it was shown~\cite{M-wc,bowles} that any experimental setup with a non-zero detection efficiency can be used to prove nonclassicality. Experimental realization of protocols based on these ideas has been shown recently~\cite{ex1,ex2}. However, lack of shared randomness is a strong assumption, difficult to justify in practice. On the other hand, is seems to be relatively easy to check the classical channel capacity. It suffices to establish the type and range of parameters accessible to sender's device and the classical capacity is bounded by the number of orthogonal states which can be prepared. This can be done by a close inspection of the hardware. Such an inspection is not trivial but overwhelmingly easier than full characterization of the device.    

 Here we propose a method for modification of the existing communication based protocols, which enables to relax this assumption. The modified protocols retain their potential to provide conclusive results for any positive detection efficiency and do not require any additional constraints. This leads to a huge increase in reliability of the tests of nonclassicality.

Note that the assumption we need for our proposal to work is limited to classical capacity. This is important, as the quantum capacity is by far more difficult to be verified. The aim of the proposed protocol is to show an experimentally feasible way of testing that the transmitted system is quantum, no matter what is its dimension.

The idea behind the method is quite simple. Consider a standard nonclassicality test involving one way communication between two parties. We assume this test cannot be passed using classical resources with detectors of arbitrarily low efficiency, if the parties are \textbf{not} using shared randomness. Now, we want to provide a way to remain the test to be conclusive even if the parties \textbf{use} shared randomness.

To this end, suppose we run this test with two pairs of parties performing it in parallel. However, in every round of the test, we choose randomly the pairing between senders and receivers. If the devices rely on shared randomness in their behavior then both senders should be correlated with both receivers as they do not know to whom they are sending their messages. Yet, if this is the case, there might appear correlations in the outputs of receivers, which should not be there. If we penalize such correlations, we might end up in a situation where using shared randomness does not give any advantage.

We note that an access to a source of random variables (possibly public) which are not correlated senders and receivers devices is crucial for working of the protocol. Within the paper we use such a randomness as a resource to create an experimentally feasible setup resistant to detection efficiency loophole. This seems to be reasonable as, unless a paranoid level of certainty is assumed, it is impossible for an eavesdropper e.g. to correlate with all of publicly available sources of randomness. On the other hand the threat that devices exploit the detection efficiency loophole is quite real.

This is, obviously, only the intuitive reasoning that leads us to introducing our method. In the rest of the paper, we will show that the method indeed works for the simplest example of a communication protocol and how it can be applied in practice. We will also report the realization of a proof of principle experiment for our test of nonclassicality. We believe that this method could be applied as easily to even complex protocols, however, their study is beyond the scope of the paper. Neither do we consider the straightforward application of our method in cryptography as our aim here is a proof-of-principle.

\section{Semi Device Independent Scenario}

\begin{figure}[hbtp!]
	\centering
	\includegraphics[width=8.3cm]{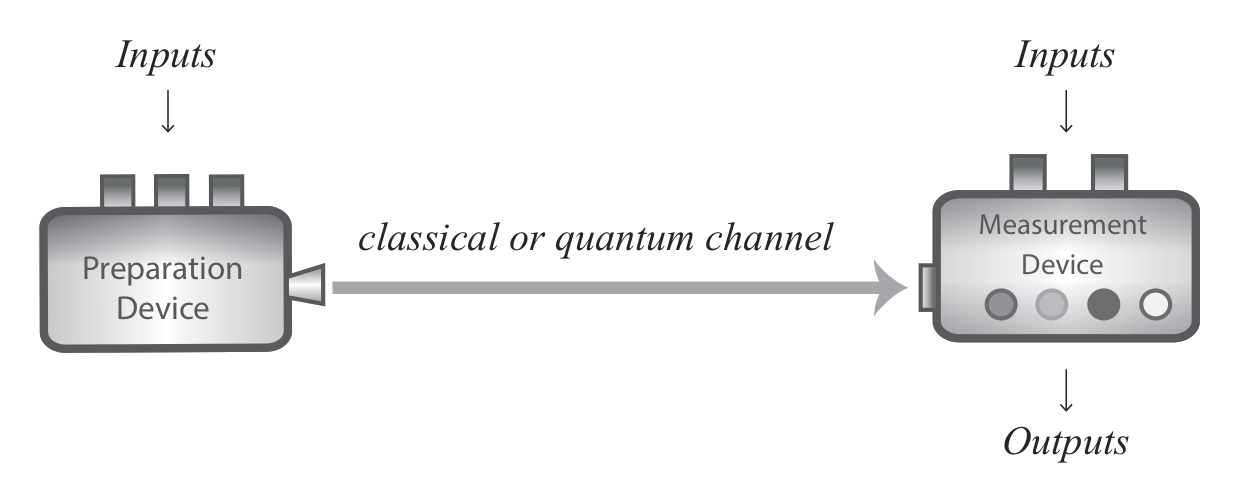}
	\caption{Semi-device independent implementation in a prepare and measure scenario. Preparation device communicates a system depending on its inputs. The message is encoded in a system with an upper bound on the dimension. The measurement device's output depends on its inputs and the received message.\label{fig1:SDI}}
\end{figure}

An ordinary SDI scenario involves two black boxes representing state preparation and measurement devices, see Fig.~\ref{fig1:SDI}. The former prepares a physical system based on the input received and communicates it, via an external channel, to the latter. The measurement device returns an output after receiving the communication and an additional classical input. There are no assumptions on the internal working of the preparation and measurement devices but we do assume an upper-bound on the information capacity of the system communicated between them. Here the classical information capacity of our system is 1 bit. We prove that the systems sent through it, necessarily, have nonclassical properties and this is realized by using a method based on a random access code. We'd like to note that these two devices are placed in a shielded laboratory, meaning they don't receive any external synchronization signals. They may share correlated classical variables but these are assumed to be uncorrelated with the inputs.

The assumption on the upper-bound of the communication capacity might seem difficult to justify. It is, however, much easier to inspect the components of the device, responsible for encoding the information in a physical system, to find its dimension then to check if every logical circuit and bit of software does what it is supposed to do. Moreover, the SDI case has been well studied. Our method would probably work just as well with other possible constraints but as it is introduced here, we have chosen to analyze it in a setting which makes it easiest.

\section{Random Access Codes}

Random Access Code (RAC) is a communication complexity problem. In its simplest case, the sender (Alice) is given an input $a$ consisting of two bits: $a_0$ and $a_1$. She is only allowed a single use of a communication channel which transmits systems of one classical bit capacity to send a message to the receiver (Bob). This means that the communicated system is a single bit (qubit) of information in the classical (quantum) case. The receiver, apart from the message, also receives an input $y\in\{0,1\}$ and his task is to return one of the bits that the sender received specified by this input, i.e. $a_y$. If we denote the output of the receiver by $b$ then the success of the task is measured by the probability $P(b = a_y|a,y)$. 

The use of RACs in quantum information was mentioned already in~\cite{Wiesner} and the interest in them has been present ever since~\cite{rac1,rac2}. In fact most of SDI protocols are based on RAC~\cite{SDI,Li1,Li2} and depending on the targeted application, different figures of merit can be used, e.g.: average success probability $S_{av}=\frac{1}{8}\sum_{a,y}P(b = a_y|a,y)$~\cite{SDI,Li1,Li2}; worst case $S_{wc}=\min_{a,y}P(b = a_y|a,y)$~\cite{rac1}; or even the whole set $\{P(b = a_y|a,y)\}_{a,y}$~\cite{ex1}. Regardless of the chosen figure of merit, quantum communication allows us to reach values, that are not possible for classical resources, thereby making RACs a good choice for a test of nonclassicality. For example, the maximum average success probability when 1 bit is communicated is 0.75 while with 1 qubit it can reach 0.85.

Like other tests of nonclassicality, RACs are also vulnerable to the detection efficiency loophole~\cite{Pearle}, which leaves them inconclusive if the detectors frequently fail to register incoming particles. The critical detection efficiency depends on the particularities of the test: the choice of the figure of merit and the corresponding treatment of the experimental rounds with no particle registered. In our work, we choose the most generic approach for the latter case where we make our devices return a random number whenever a detector fails to register a particle. This artificially makes the effective efficiency to be 100\% as we observe an outcome in every round. This does however lead to a decrease in the maximal success probability for the quantum case. For a given maximum $Q$ of the quantum theory, the detectors registering particles in $\eta$ fraction of the rounds can yield the largest success probability
\be
	\label{qeta}
	Q_{\eta}=\eta Q +(1-\eta)\frac{1}{2},
\ee
tending to $\frac{1}{2}$ as $\eta \to 0$. The classical value does not increase allowing to find the $\eta$ required for the quantum theory to be able to provide an advantage. However, the critical $\eta$ is usually very high and extremely difficult to reach in practice. Our goal is to propose a test with $\eta$ arbitrarily close to 0. From~\eqref{qeta} we see that it is possible if we can design a test with the maximal classical value reduced to $1/2$. This is the bound for the worst case success probability if no shared randomness is available to the communicating parties~\cite{M-wc}. As mentioned, we need tasks with such a property as a starting point for our method. This, together with its simplicity makes RAC a perfect candidate.

\section{Probability Polytopes}
The considerations above are specific to RACs. However they can be generalized to different tests of nonclassicality using the framework of probability polytopes~\cite{dimwit}. Consider a set of conditional probability distributions $p(\vec{x}|\vec{y})$ arising in an experiment, where $\vec{x}$ denotes the set of outcomes of all the devices in a single round of the experiment and $\vec{y}$ the inputs. Let $\vec{p}$ be a vector whose elements correspond to conditional probabilities for every combination of $\vec{x}$ and $\vec{y}$. $\vec{p}$ can characterize the behavior of any devices employing any strategy. While testing these devices, e.g. if they violate Bell inequalities, we often look at the linear combinations of probabilities $p(\vec{x}|\vec{y})$. Such tests can also be represented by a vector, $\vec{t}$, living in the same space as $\vec{p}$. The quantitative outcome of the test is the scalar product $\vec{t} \cdot \vec{p}$.

The average success probability of a RAC, $S_{av}=\frac{1}{8}\sum_{a,y}P(b = a_y|a,y)$, is an example of such a test. The vector $\vec{t}_{RAC}$ corresponding to it has 16 elements because there are that many combinations of $a,b$ and $y$. These elements are equal to $\frac{1}{8}$ whenever the corresponding variables satisfy $b = a_y$ and 0 otherwise. When a single bit is communicated, then for every possible classical protocol we have $S_{av}=\vec{t}_{RAC}\cdot\vec{p}_{cl}\leq 0.75$.

For classical devices the set of allowed probability distributions $\vec{p}_{cl}$ is a polytope. Its vertices correspond to deterministic strategies. Every other strategy can be considered as a convex combination of deterministic ones and corresponds to a point inside this polytope. The set of quantum distributions is much harder to characterise and, instead of faces, usually is bounded by smooth curved surfaces. This set is, typically, strictly larger than the classical polytope but their dimensions are usually the same. However it is not always true~\cite{Joh} and these are the cases that we find the most interesting. Whenever this happens, there exists a vector $\vec{t}_0$ perpendicular to the subspace spanned by classical probabilities, i.e. $\forall_{\vec{p}_{cl}} \quad \vec{t}_0\cdot\vec{p}_{cl}=0$, and there exist a quantum probability distribution $\vec{p}_{qm}$ such that $\vec{t}_0\cdot\vec{p}_{qm}=Q\neq 0$.

Consider again the case with an experiment reaching $\vec{p}_{qm}$ if we had perfect detectors. If no particles are registered, we again assign a random outcome. Since returning a random outcome is also a classical strategy, the vector corresponding to the probability distribution observed in the experiment is $\vec{p}_{\eta}=\eta \vec{p}_{qm}+(1-\eta)\vec{p}_{cl}$ and for $\eta > 0$ the outcome of the test will be equal to $\eta^Q > 0$, since by assumption $\vec{p}_{qm} > 0 $ and $\vec{p}_{cl} = 0$. Therefore, it pays off to look for experimental setups which have a different dimensions of the corresponding quantum and classical sets.

We will see that the two approaches: looking for a game with classical success probability $\frac{1}{2}$ and for setups with different classical and quantum set dimensions, are complementary. The former provides us with intuitive candidate in the form of a parallel RAC that we describe in the next section. However, what we need from it is only the structure of the experiment, i.e.: number of parties and their inputs and outputs, communication paths, and the quantum protocol. The proof that the experiment with the given structure provides us a robust test given by the latter and in the test itself no reference to RACs is necessary.

\section{Parallel Random Access Coding}

We define parallel random access code as a task in which two pairs of senders and receivers perform RAC in parallel but in each round they are paired randomly. Additionally, the devices do not have access to information about the pairing in any given round. This situation is illustrated in Fig. 1. The inputs to preparation device $P^i$ is denoted by $a^i$, and by $y^i$ for the measurement devices, $M^i$, $i=0,1$. The choice of pairing is denoted by a random variable $x$. If $x=0$, information from $P^0$ goes to $M^0$ and from $P^1$ to $M^1$. Whereas for $x=1$, $P^0$ communicates with $M^1$ and $P^1$ with $M^0$. We state that the $n$th receiver (i.e. measuring device) is successful if $b^{n} = a^{n \oplus x}_{y^n}$, where $n$ is either $0$ or $1$.

\begin{figure}[hbtp!]
	\centering
	\includegraphics[width=8.1cm]{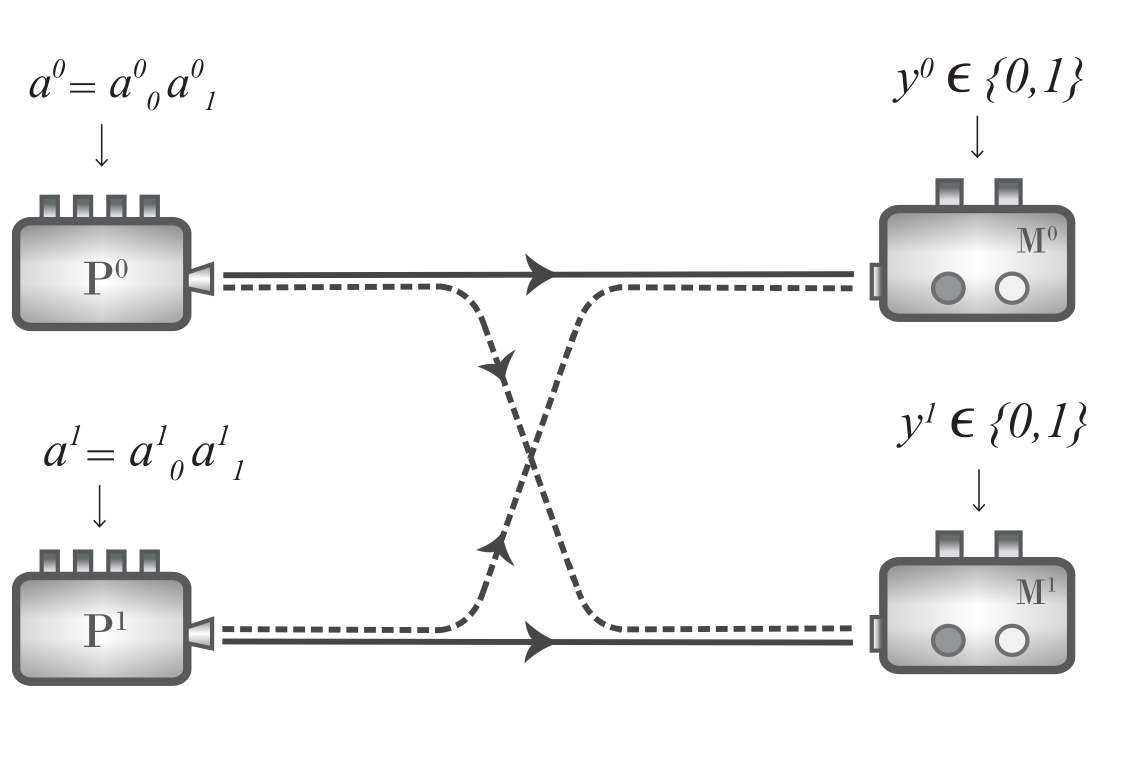}
	\caption{Parallel RAC scenario consisting of two preparation devices, $P^0$ and $P^1$, and two measurement devices, $M^0$ and $M^1$. The pairing choice between the senders and receivers devices depends on the value of random variable $x$. Here, the solid line corresponds to the $x=0$, and the dotted line to $x=1$.}
\end{figure}

There exist $2^{16} = 65536$ different deterministic strategies. Because of symmetries between some of them (e.g. joint negation of messages of both Alices connected with negation of the reaction of both Bobs), these strategies relates to $30496$ different point in conditional probability space,
\be
	P(b^{out}_0, b^{out}_1 | a_{0,0}, a_{0,1}, a_{1,0}, a_{1,1}, c, b_0, b_1),
\ee
where $b^{out}_i$ refers to the outcome of $i$-th Bob, $a_{i,j}$ to $j$-th bit of $i$-th Alice, $c$ determines which Alice is sending to which Bob, and $b_i$ is the input of $i$-th Bob.

Since there are $4$ different possible outcomes, $(b^{out}_0, b^{out}_1)$, and $2^7=128$ different possible inputs $(a_{0,0}, a_{0,1}, a_{1,0}, a_{1,1}, c, b_0, b_1)$, the points in this probability space can be represented as vectors forming a simplex in $4 \times 128 = 512$ dimensional linear space. It reveals that this simplex is embeded in $125$ dimensional subspace (i.e. there are only $125$ linearly independent vectors among $30496$ points). Let $P^{sub}$ be the projector onto that subspace.

Let $P^{max}$ be the vector in $512$ space related to the optimal quantum strategies maximizing the probability of correct guesses of both Bobs. If we project the vector $P^{max}$ on $P^{sub}$ then we obtain a Bell operator $B^{2s2r}_{1,1}$ given by
\begin{widetext}	
	\be
		B^{2s2r}_{k,l} \equiv \sum_{\substack{a_{0,0}, a_{0,1} \\ a_{1,0}, a_{1,1} \\ b_0, b_1} = 0, 1} \delta_{b_0 \oplus b_1, k} \delta_{a_{0,0} \oplus a_{0,1} \oplus a_{1,0} \oplus a_{1,1}, l} \times \left[ \sum_{b^{out}_0, b^{out}_1 = 0, 1} (-1)^{s_0 + s_1} P(b^{out}_0, b^{out}_1 | a_{0,0}, a_{0,1}, a_{1,0}, a_{1,1}, c, b_0, b_1) \right],
	\ee
\end{widetext}
where $s_i$ is the binary indicator success of $i$-th Bob. Let $p = \frac{1}{2} \left(1 + \frac{\sqrt(2)}{2} \right)$ be the maximal probability of success of a $2$ to $1$ QRAC. It is easy to see, that for each non-zero term in the outer summation the maximum value is $p^2 + (1-p)^2 - 2 p (1-p) = 0.5$, and there are $32$ such a terms, giving Tsirelon bound $16$. The values obtained in the experiment are $B^{2s2r}_{0,0} = 0.1797$, $B^{2s2r}_{0,1} = 0.1792$, $B^{2s2r}_{1,0} = 0.1728$ and $B^{2s2r}_{1,1} = 0.1721$.

In deterministic model each Alice receives $4$-dimensional input (a setting) and returns a binary output, so each Alice has $2^4=16$ deterministic strategies. Similarly each Bob get a $4$-dimensional input (message and setting), and outputs a binary result. Thus, in our setting, there exist $2^{16}$ deterministic strategies. They are spanning a $125$ dimensional subspace. Now we need to choose a vector corresponding to the test that we are going to perform. We choose it to be parallel to the line connecting the point $\vec{p}_{qm}$ corresponding to the probabilities that would arise in the perfect quantum experimental realization of RAC and its projection onto the classical subspace. It gives rise to the following figure of merit
\be
	\ba
		T = \sum_{a_0^0\oplus a_0^1\oplus a_1^0\oplus a_1^1=1} & \sum_{y^0\neq y^1} \sum_{b^0,b^1,x}\\
		& s_0s_1 p(b^0,b^1|a^0,a^1,y^0,y^1,x),
	\ea
\ee
where $s_0$ and $s_1$ are success indicators of the two receives. $s_n=1$ when $b^{n} = a^{n \oplus x}_{y^n}$ and -1 otherwise. Note there is no obvious relation between T and average or worst case success probability in RAC. This code was only a step, necessary to find the number of inputs and outputs for each of the parties, defining the space and point $\vec{p}_{qm}$. However, it is still quite straightforward to turn $T$ into a game with classical average success probability of $\frac{1}{2}$ simply by changing each coefficient in $T$ from -1 to 0, i.e. we replace $s_0 s_1$ with Kronecker's delta $\delta_{s_0,s_1}$ and get
\be
	\ba
		T' \equiv \sum_{a_0^0\oplus a_0^1\oplus a_1^0\oplus a_1^1=1} & \sum_{y^0\neq y^1} \sum_{b^0,b^1,x} \\
		& \delta_{s_0,s_1} p(b^0,b^1|a^0,a^1,y^0,y^1,x).
	\ea
\ee
$T'$ describes average success probability in a game in which success is defined by having $s_0=s_1$. If $T=0$, then $T'=\frac{1}{2}$, therefore it is a game with classical average success probability $\frac{1}{2}$.

One way to view $T'$ is as a parallel RAC with a promise on the distribution of the inputs, expressed by the fact that the sums in $T'$ do not include every possible combination of input variables. Another important aspect is that the parties win a round of the game not if they successfully guess the desired bit as in a typical RAC but when their successes are correlated, i.e. they either both guess correctly or both fail. However it's easy to see that using the same states as measurements as in a standard RAC the parties can beat the classical bound. It suffices to notice that the optimal strategy in that protocol gives success probability $p_s=\frac{1}{2}+\frac{1}{2\sqrt{2}}$ for every input. If the same states and measurements are used in this game one will obtain $T'=p_s^2+(1-p_s)^2=\frac{3}{4}$.

From now on, one can stop thinking about RACs and consider only T. Its classical value is 0 and the theoretical maximal quantum value is 16 corresponding to the point $\vec{p}_{qm}$. Therefore, our result is a general property of a probability space corresponding to a parallel test with a certain number of inputs and outcomes rather than of RACs.

We leave as an open question whether the same (i.e. that classical polytope has lower dimension than the quantum region) holds for spaces corresponding to tests with different than binary numbers of inputs and outcomes. This seems plausible and it is this conjecture which is behind our optimism regarding the possibility of applying the same method in different scenarios. Now, we describe our experimental realization to obtain an experimentally measured value of T.

The experiment requires three assumptions:
\begin{itemize}
	\item The system that leaves the lab of each sender has information encoded in the Hilbert space dimension of 2.
	\item The devices don't have the access to the information about the party they are communicating with in the given round of the experiment.
	\item The inputs of all the parties are chosen at random.
\end{itemize}

In our experiment we measured each combination of settings for 20 seconds before moving to the other, therefore the final assumption was not satisfied. This is because it was intended as a proof-of-principle and choosing random settings for each round would considerably extend the, already long, measurement time. In practical applications based on this setup that assumption cannot be omitted.

\section{Experimental Demonstration}

\begin{figure}[h]
	\centerline{\includegraphics[width=1\columnwidth]{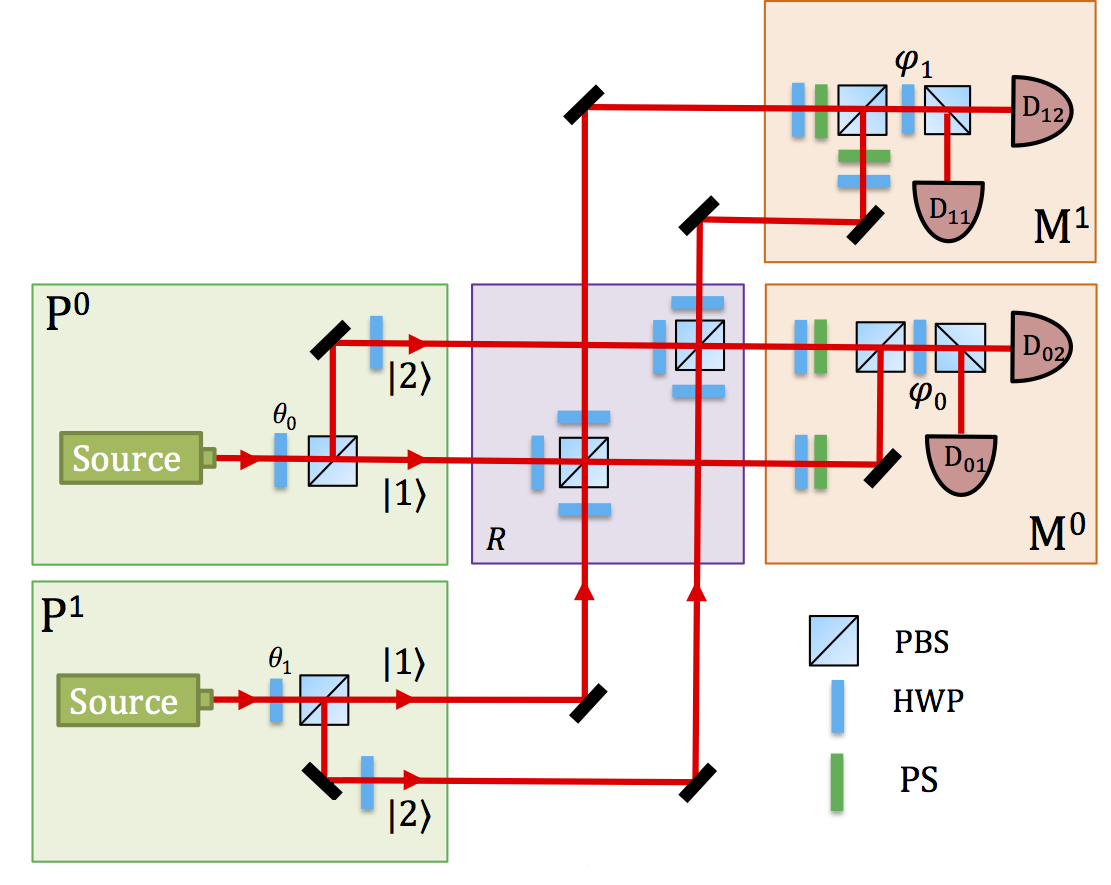}}
	\caption{Experimental setup. Information is encoded into two spatial-photonic modes. The preparation devices $P_i$ ($i = 0,1$) consist of a heralded single photon source that emits horizontally polarized photons which, after passing through a half-wave plate (HWP) oriented at an angle $\theta_{i}$, can be prepared in the required protocol states. The measurement devices $M_i$ consist of a HWP oriented at an angle $\varphi_{i}$, polarization beam splitter (PBS), single photon detectors $D_{ij}$. The pairing of the preparation and measurement devices is performed with help of adjustable HWP plates and PBSs. }
	\label{Fig4}
\end{figure}

Our experimental realization is shown in Fig.~\ref{Fig4}. Any qubit state can be prepared by a suitably oriented half-wave plate, HWP($\theta_0$) and HWP($\theta_1$) for $P_0$ and $P_1$ respectively. We used a heralded single photon source from spontaneous parametric down conversion process at 780 nm.
For the four experimentally prepared qubit states, the HWP settings for both $\theta_0$ and $\theta_1$ correspond to $0^\circ, 45^\circ$ and $\pm22.5^\circ$. We have also added an extra HWP in each device to assure the same polarization in both paths.

The choice of the two communication paths is made in a region $R$, which consists of two PBSs and six HWPs, and by properly adjusting the HWP orientation angles, we choose the pairing of the devices. Note that we do not consider the region $R$ a device for the purposes of our analysis. Another option would be simply to connect senders and receivers by fibers randomly each round. This would leave $R$ empty but greatly increase the duration of the experiment.

The measurement devices consist of an interferometric setup, one adjustable HWP($\varphi_0$) and HWP($\varphi_1$) for $M_0$ and $M_1$ respectively along with two polarization beam splitters (PBS) and two single-photon detectors ($D_{ij}$, $i=0,1; j=1,2$) for each device $M_j$. The success probabilities are estimated from the number of detections in the detectors $D_{ij}$, after properly adjusting the orientation $\varphi_j$ of the half wave plate in each of the measurement devices $M_0$ and $M_1$. In our experiment, the two used HWP settings are $11.25^\circ$ and $78.75^\circ$ respectively.

Considering the parallel RAC scenario in figure 2, when $x=0$, the state prepared in $P^0$ ($P^1$) is measured in the receiver $M^0$ ($M^1$), whereas for x=1, the state prepared in $P^0$ ($P^1$) is measured in the receiver $M^1$ ($M^0$). In this proof of principle experimental demonstration, the values of HWP($\theta_0$) and HWP($\theta_1$) on the preparation sides and HWP($\varphi_0$) and HWP($\varphi_1$) on the measurement sides were preselected rather than a randomized selection. However, a randomized selection between the settings choice can be implemented by mounting all the corresponding wave plates on motorized stages.

Our single-photon detectors ($D_{ij}$, $i=0,1; j=1,2$) were fiber coupled silicon avalanche photodiodes (APD) with effective detection efficiency $\eta_d = 0.55$ (with 0.85 fiber coupling and 0.65 for APD efficiency) and dark counts rate $R_d \simeq 300$ Hz.
The detectors $D_{ij}$ produced output transistor-transistor logic signals of 4.1 V (with duration of $41$ ns). The dead time of the detectors was $50$ ns. All single counts were registered using multi-channel coincidence logic with a time window of $1.7$ ns.

For each choice of settings  $a^0,a^1,y^0,y^1,x$, we registered an average of 222528 clicks, where 3104 we have the four-photon coincidence, meaning that both photon sources in $P^0$ and $P^1$ were heralded. Only the runs that we have both photons heralded together  contribute for the value of $T$, for all other (when one or no photon is heralded) $\sum_{b^0,b^1}s_0s_1 p(b^0,b^1|a^0,a^1,y^0,y^1,x)=0$, where we considered that for these rounds a random value was assigned to the outcomes. If we consider only the run when we measured four-photon coincidence and fair-sampling, we have $T=14.74\pm1.56$. Now, considering the detector efficiency and all runs,  we obtain $T=3.16\pm0.33~\times~10^{-3}$, witnessing the system  nonclassicality. 


\section{Conclusion}

The nonclassicality test, that we have introduced and experimentally demonstrated, allows for arbitrarily low detection efficiency without invoking extra assumptions such as independence of the devices. This opens up a whole new possibility for constructing semi-device independent protocols based on this test, which can be easily realized with today's technology.

\begin{acknowledgments}
This work is supported by FNP grant First TEAM, NCN grant 2014/14/E/ST2/00020, Swedish Research Council (VR), ADOPT and DS Programs of the Faculty of Electronics, Telecommunications and Informatics, Gda\'nsk University of Technology. The ’International Centre for Theory of Quantum Technologies’ project is carried out within the International Research Agendas Programme of the Foundation for Polish Science co-financed by the European Union from the funds of the Smart Growth Operational Programme, axis IV: Increasing the research potential (Measure 4.3).
\end{acknowledgments}

\end{document}